\documentclass[12pt]{article}
\pdfoutput=1

\usepackage{amsmath,amssymb,amscd}
\usepackage{listings}
\usepackage{caption}
\usepackage{dsfont}
\usepackage{slashed}
\usepackage{color}

\usepackage[pdftex]{graphicx}
\usepackage{epstopdf}
\usepackage{subfigure}
\usepackage{epsfig}
\usepackage{listings}
\usepackage{caption}
\usepackage{cite}

\usepackage{multirow}

\newcommand{\bea}{\begin{align}}
\newcommand{\eea}{\end{align}}
\newcommand{\beq}{\begin{equation}}
\newcommand{\eeq}{\end{equation}}
\newcommand{\nbea}{\begin{align*}}
\newcommand{\neea}{\end{align*}}
\newcommand{\nbeq}{\begin{equation*}}
\newcommand{\neeq}{\end{equation*}}
\newcommand{\bear}{\begin{eqnarray}}  
\newcommand{\eear}{\end{eqnarray}}  

\numberwithin{equation}{section}

\begin{document}

\def\thefootnote{\fnsymbol{footnote}}

\begin{flushright}
{\tt KCL-PH-TH/2026-27, CERN-TH-2026-281}
\end{flushright}

\vspace{0.7cm}
\begin{center}
{\bf {\Large The search for the Higgs Boson \\ \vspace{2mm}
and the EBH mechanism}}
\end{center}

\medskip

\begin{center}{\large
{\bf John~Ellis}}
\end{center}

\begin{center}
{Theoretical Particle Physics and Cosmology Group, Department of
  Physics, King's~College~London, London WC2R 2LS, United Kingdom;\\
Theoretical Physics Division, CERN, CH-1211 Geneva 23,
  Switzerland}
  \end{center}

\vspace{0.7cm}

\centerline{\bf ABSTRACT}
The searches for the Higgs boson at LEP and the Tevatron between 1980 and 2000 are set in the context of the underlying theory of spontaneous symmetry breaking and early phenomenological studies. The evolution of direct searches at LEP and the Tevatron are reviewed, as well as indirect constraints from precision electroweak data. Finally, the state of play in 2000 is summarised, together with the perspectives it offered for later searches.~~\\
\noindent  

~~\\
\begin{center}
Contribution to the 4th International Symposium on the History of Particle Physics \\
covering the period between 1980 and 2000, held at CERN in November 2025
\end{center}

\vspace{0.7cm}
\begin{flushleft}
November 2026
\end{flushleft}

\newpage

\section{Prehistory}

For a physicist, the vacuum is not simply empty space, but the lowest-energy state allowed by the underlying physical laws. The vacuum is determined by  classical and quantum-mechanical processes, and need not possess all the symmetries of the underlying equations
of the physical system, a phenomenon known as `spontaneous' symmetry breaking,
or `hidden' symmetry.

Spontaneous symmetry breaking came to prominence in
the theory of superconductivity proposed by Bardeen, Cooper and Schrieffer~\cite{BCS}.
According to this theory, the photon acquires an effective mass~\cite{Anderson} through interactions with phonons in a medium when it propagates through certain materials
at sufficiently low temperatures, as discussed earlier by Ginzburg and Landau~\cite{GL}. In free space, 
the masslessness of the photon is guaranteed by Lorentz invariance and U(1) gauge symmetry, whereas a superconductor has a well-defined rest frame, so Lorentz invariance is broken explicitly. However, the gauge
symmetry is still present, though `hidden' by the condensation of Cooper pairs of electrons~\cite{Cooper} in the lowest-energy
state (vacuum). It was explicitly shown by Anderson~\cite{Anderson} how the interactions 
with the photon of the Cooper pairs inside a superconductor cause the former to acquire an effective mass.

The idea of spontaneous symmetry breaking was introduced into particle physics by Nambu~\cite{Nambu} in 1960. 
He suggested that the small mass and low-energy interactions of pions could be understood as a reflection of a
spontaneously-broken global chiral symmetry, which would have been exact if the up and down quarks were massless. His
suggestion was that light quarks condense in the vacuum, much like the Cooper pairs of superconductivity.
When this happens, the `hidden' chiral symmetry causes the pions to become massless Goldstone bosons~\cite{Goldstone,GSW}, and fixes their low-energy couplings 
to protons, neutrons and each other.

However, the appearance of a massless boson may be avoided 
if it is a gauge symmetry that is broken,
as in the non-relativistic case of superconductivity~\cite{Anderson}.
Anderson conjectured that it should be possible to extend this
mechanism to the relativistic case, as did Klein and Lee~\cite{KL},
but it was argued by Gilbert~\cite{Gilbert} that Lorentz invariance
would forbid this.

\section{And then there was Higgs}

Spontaneous breaking of gauge symmetry was introduced into particle
physics in 1964 by Englert and Brout~\cite{EB}, followed independently
by Higgs~\cite{H1,H2}, and subsequently by Guralnik, Hagen and
Kibble~\cite{GHK}. They demonstrated how one could dispose
simultaneously of two unwanted massless bosons: a spinless Nambu-Goldstone
boson appearing when the vacuum breaks the underlying global symmetry~\cite{Goldstone,GSW} and a gauge boson of an exact local symmetry, by combining them
into a single massive vector boson in a fully relativistic theory.
The two polarisation states of a massless vector boson are combined
with the single degree of freedom of a spin-zero particle to yield the
three degrees of freedom of a massive spin-one particle $V$ with mass:
\beq m_V = g_V \, \frac{v}{\sqrt{2}} \, ,\label{Vmass}\eeq
\noindent
where $g_V$ is the corresponding gauge coupling constant and $v$ is the symmetry-breaking vacuum expectation value (vev). This is the EBH mechanism, and a key prediction is that the vector boson mass is proportional to the gauge coupling.

Englert and Brout~\cite{EB} considered explicitly a non-Abelian Yang-Mills theory, assumed the formation of a vev for of a non-singlet scalar field, and used a diagrammatic approach to demonstrate mass
generation for the gauge field. The first of two papers written by Higgs~\cite{H1} in ignorance of the Englert-Brout paper demonstrated that gauge symmetry
provides a loophole in the `no-go' theorem of Gilbert mentioned above, and his second paper~\cite{H2} exploited this loophole
to demonstrate mass generation explicitly in the Abelian case. The subsequent paper 
by Guralnik, Hagen and Kibble~\cite{GHK} referred explicitly to the Englert-Brout and Higgs papers, and
also demonstrated mass generation in the Abelian case.

The second paper by Higgs~\cite{H2} is the only one of the 1964 papers
to mention explicitly [his equation (2b)] the existence of a massive
scalar particle associated with the curvature of the effective
potential that determines the vev $v$ of the charged field:
\beq m_H = \sqrt{2\lambda} \, v \, .\label{Hmass}\eeq 
\noindent
Englert and Brout~\cite{EB} do not discuss the spectrum of physical
scalars, whilst Guralnik, Hagen and Kibble~\cite{GHK} mention a
massless scalar that is decoupled from the massive excitations in their
model.

Also worthy of note is a remarkable paper written in ignorance of these papers by Migdal and Polyakov in 1965~\cite{MP},
while they were still students~\footnote{It was finally published in 1966
after a substantial delay caused by the scepticism of Soviet academicians.}, in which they
discuss partial spontaneous symmetry breaking in the non-Abelian case.
The year 1966 saw a further important paper by Higgs~\cite{H3},
in which he discussed in detail the formulation of the spontaneously-broken Abelian theory.
In particular, he derived explicitly the Feynman rules for processes involving what has come to be
known as the massive Higgs boson, discussing its decay into 2 massive vector
bosons, as well as vector-scalar and scalar-scalar scattering. Another important paper by Kibble~\cite{Kibble}
discussed in detail the non-Abelian case, including partial spontaneous symmetry
breaking, as well as mentioning the appearance of massive scalar bosons
\`a la Higgs.

The next important step was the incorporation by
Weinberg~\cite{Weinberg} and by Salam~\cite{Salam} of non-Abelian
spontaneous symmetry breaking into Glashow's~\cite{Glashow} unified
SU(2)$\times$U(1) model of the weak and electromagnetic
interactions. Weinberg's paper was the first to observe that the
scalar field vev could also give masses to fundamental fermions $f$ that
are proportional to their couplings to the Higgs boson:
\beq m_f = g_{f\bar f H} \, v \, ,\label{fmass}\eeq
another manifestion of the EBH mechanism. 

These seminal papers on the spontaneous breaking of gauge symmetries and
electroweak unification were largely ignored by the particle physics
community until the renormalisability of spontaneously-broken gauge theories was demonstrated by
't Hooft and Veltman~\cite{tHV}. These ideas then joined rapidly the theoretical mainstream, thanks
in particular to a series of influential papers by B.~W.~Lee and collaborators~\cite{LZJ,FLS}.

\section{A Phenomenological Profile of the Higgs Boson}

B.~W.~Lee also carries much of the responsibility for calling the
Higgs boson the Higgs boson, mentioning repeatedly `Higgs scalar
fields' in a review talk at the International Conference on
High-Energy Physics in 1972~\cite{Lee}. However, despite its theoretical respectability, in the early 1970s
there were only a few suggestions how to constrain or exclude the
possible existence of a physical Higgs boson. One paper considered the
possible effect of Higgs exchange on neutron- and deuteron-electron
scattering and derived a lower bound $m_H > 0.6$~MeV~\cite{nue}, and
another constrained Higgs emission from neutron stars, yielding the
lower bound $m_H > 0.7$~MeV~\cite{SS}.  There was also a theoretical
discussion of possible Higgs production in $0^+ \to 0^+$ nuclear
transitions~\cite{RSW}, and its non-observation in excited $^{16}$O
and $^4$He decays led to the Higgs being excluded in the mass range
$1.03~{\rm MeV} < m_H < 18.3$~MeV~\cite{KWB}. In parallel, data on
neutron-nucleus scattering were used to constrain $m_H >
15$~MeV~\cite{BE}.  The two latter were the strongest limits obtained
in this period.

This was the context in which we embarked in 1975 on the first systematic study of possible Higgs phenomenology~\cite{EGN}.
Neutral currents had been discovered~\cite{NC}, as had the J/$\psi$ particle~\cite{Jpsi}, which was thought to be charmonium, though a few
doubts remained as the discovery of open charm still lay in the future. The search for the intermediate vector
bosons $W^\pm$ and $Z^0$ was appearing on the experimental agenda, but the CERN ${\bar p}p$
collider that was to discover them had not yet been proposed. However, it seemed to us that the clinching
test of the spontaneous symmetry-breaking paradigm underlying the Standard Model would be discovering
the Higgs boson. Its discovery would provide the capstone of the theoretical construction of the Standard Model.

\begin{figure}[htb]
\begin{center}
\resizebox{9cm}{!}{\includegraphics{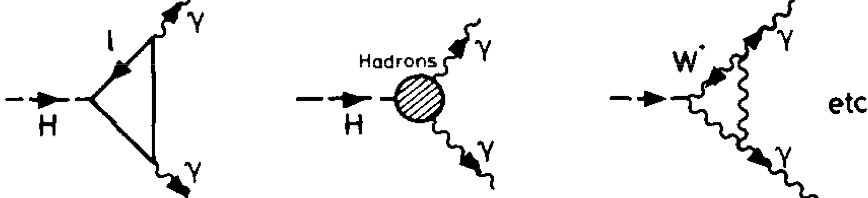}}
\end{center}
\caption{\it One-loop diagrams contributing to $H \to \gamma \gamma$ decay.  }
\label{fig:Hgammagamma}
\end{figure}

To this end, we considered the decay modes of the Higgs boson if it weighed up to 100~GeV, recalling that its tree-level couplings to other particles would be proportional to their masses (\ref{Vmass}, \ref{fmass}), and calculating for
the first time the loop-induced Higgs decays to photon pairs~\cite{EGN}, see Fig.~\ref{fig:Hgammagamma}.
The dominant mechanism for this decay is an
anomalous triangle diagram with a $W^\pm$ loop, and there are also subdominant diagrams with
massive quarks~\footnote{We did not calculate these: the $t$ and $b$ had not been discovered at that time.}~\cite{Djouadi}:
\beq
\Gamma(H \to \gamma \gamma) \; = \; \frac{G_\mu \alpha^2 m_H^3}{128 \sqrt{2} \pi^3}
\left| A_1 (r_W) + \sum_f N_c Q_f^2 A_{1/2} (r_f)\right|^2 \, ,
\label{Djouadi}
\eeq
where
\beq
A_1(r) \; \equiv \; -\left[2 r^2 + 3 r + 3(2 r - 1)f(r)\right]/r^2, \; \; A_{1/2}(r) \; \equiv\;  2 \left[r + (r - 1) f(r) \right]/r^2
\eeq
and $f (r) \equiv \arcsin^2 \sqrt{r}$ for $r_{W, f} \equiv m_H^2/4m_{W, f}^2$. 

We estimated the Higgs decay branching ratios and cross
sections for many different mechanisms for producing the Higgs boson, intending to cover the full allowed mass range
from ${\cal O}(15)$~MeV upwards. In addition to considering the production of a relatively light Higgs boson in
hadron decays and interactions, we also considered production in $e^+ e^-$ collisions, including $Z^0$ decays and Higgs-strahlung 
processes such as $e^+ e^- \to Z^0 + H$~\cite{EGN}~\footnote{These processes were also considered in~\cite{EG76,Bj,IK} and ~\cite{LQT}.}.

Back in 1975, the existence of the Higgs boson still seemed speculative, and the likelihood of a definitive search for the Higgs boson seemed somewhat remote. That was
why, somewhat tongue-in-cheek, we closed our paper~\cite{EGN} with the following cautionary words: {\it ``We should
perhaps finish our paper with an apology and a caution. We apologise to experimentalists for having no idea what is the
mass of the Higgs boson, ..., and for not being sure of its couplings to other particles, except that they are
probably all very small. For these reasons, we do not want to encourage big experimental searches for the Higgs boson,
but we do feel that people doing experiments vulnerable to the Higgs boson should know how it may turn up."}

\section{How heavy could the Higgs boson be?}

\begin{figure}[htb]
\begin{center}
\resizebox{7cm}{!}{\includegraphics{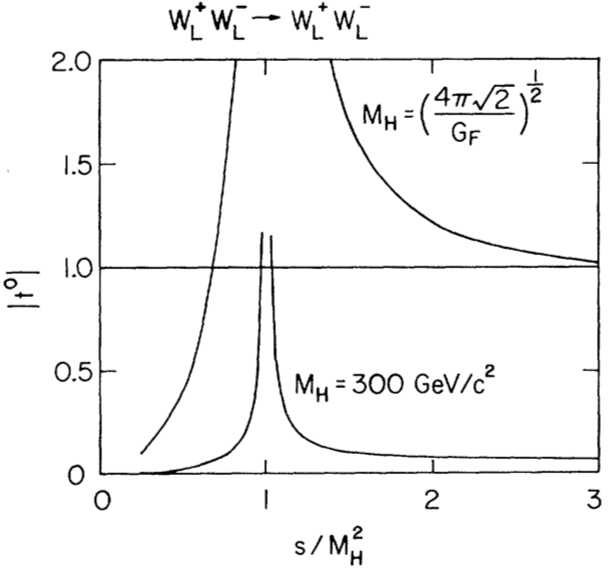}}
\hspace{1cm}
\resizebox{7cm}{!}{\includegraphics{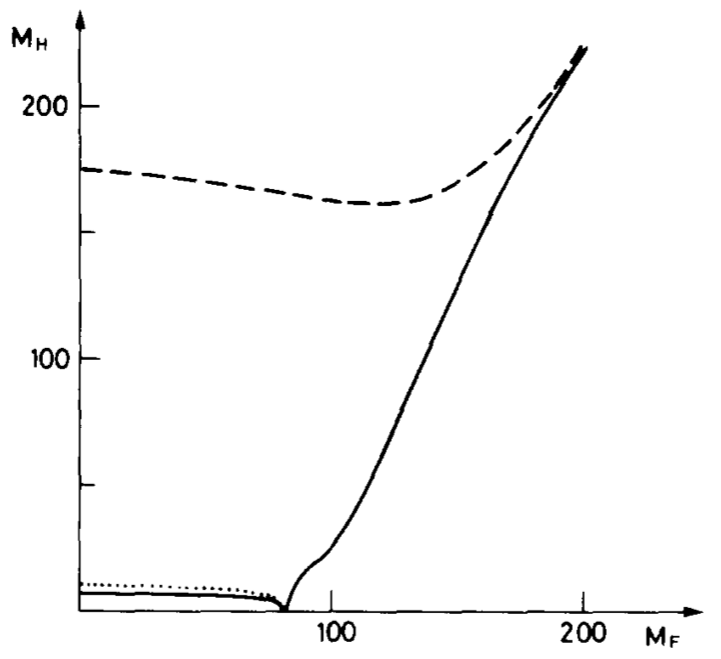}}
\end{center}
\caption{\it Left panel: Energy dependence of the $J=0$
partial-wave amplitude for the elastic scattering of longitudinally-polarised $W$ bosons for $M_H = (4 \pi \sqrt{2}/G_F)$ and $300$~GeV~\cite{LQT}. Right panel: Bounds in a grand unified theory on the mass of the Higgs boson as a function of the top quark mass~\cite{Cabibbo:1979ay}. The dashed line is the upper bound assuming perurbativity, and the solid line is obtained by requiring that the quartic Higgs coupling stays positive, ensuring the stability of the electroweak vacuum.
}
\label{fig:heavyH}
\end{figure}

In those early days, there was very little theoretical guidance as to
the possible mass of the Higgs boson, which is one reason why these
early studies also included very low Higgs masses. One possibility
that attracted attention was that the Higgs mass was entirely due to
quantum corrections, which would have yielded $m_H \sim 10$~GeV in the
absence of heavy fermions~\cite{CW}~\footnote{This was long before it
was recognised that the top quark weighed $> m_W$.}. At the other
end of the mass scale, the important point was made by Ben Lee, Chris Quigg and Hank Thacker, in particular, that the Higgs
self-interactions would become strong for $m_H \sim
1$~TeV~\cite{heavyH}, see Fig.~\ref{fig:heavyH}, which could be construed as an upper limit on the Higgs mass.
Another important paper by Nicola Cabibbo, Luciano Maiani, Giorgio Parisi and Roberto Petronzio~\cite{Cabibbo:1979ay} was the forerunner of all the current discussion about the instability of the electroweak vacuum. If there is a heavy fermion, to the right of the solid line in Fig.~\ref{fig:heavyH}, the electroweak vacuum would not be stable. As we know now, the masses of the top quark and the Higgs boson are very close to that solid line, quite possibly below it, and there is ongoing discussion whether there is some reason that the masses are so close to it, and how to interpret a lack of stability.

\begin{figure}[htb]
\begin{center}
\resizebox{9cm}{!}{\includegraphics{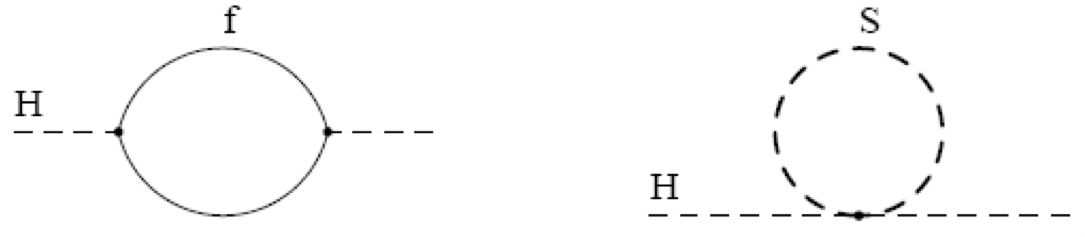}}
\end{center}
\caption{\it One-loop diagrams making quadratically-divergent corrections to the mass of the Higgs boson that cancel in a supersymmetric theory.}
\label{fig:Hloops}
\end{figure}

In parallel to this work on phenomenological aspects of the Higgs boson, some theorists were getting worried. The issue was the fact that when you calculate loop diagrams with an elementary Higgs boson, you encounter quadratic divergences. (See Fig.~\ref{fig:Hloops} for the worrisome one-loop diagrams.) One suggested mitigation was to postulate low-energy supersymmetry (see the talk by Riccardo Barbieri at this meeting), which could control the loop diagrams via  cancellations of the quadratic divergences between fermion and boson loops. 
\begin{equation}
\Delta m_H^2 \; = \; - \frac{y_f^2}{16 \pi^2} \left[2 \Lambda^2 +6 m_f^2\ln{\frac{m_f^2}{\Lambda^2}} + \dots \right], \;
\Delta m_H^2 \; = \; - \frac{\lambda_s}{16 \pi^2} \left[ \Lambda^2 -2 m_f^2\ln{\frac{m_f^2}{\Lambda^2}} + \dots \right]
\label{eq:Hloops}
\end{equation}
The terms quadratic in the ultraviolet cutoff $\Lambda$ in Eq.~(\ref{eq:Hloops}) would cancel if the fermion coupling $y_f$ and the boson coupling $\lambda_s$ obey the relation $\lambda_s = 2 y_f^2$, which holds in a supersymmetric theory. 

Another possibility for taming the quadratic divergences was to go back to the original theory of superconductivity and postulate that the Higgs boson is not an elementary particle, but in fact a composite particle made up out of constituent fermions. In such a composite Higgs model, you do not have to worry about the quadratic divergence because the Higgs boson gets dissolved into its individual constituents at high momenta. Prominent proponents of such composite scenarios included Steven Weinberg, Leonard Susskind and Savas Dimopoulos~\cite{TC}.

\section{Searches for the Higgs Boson at LEP}

Searches for the Higgs boson moved up the experimental agenda following the discoveries of the $W^\pm$ and $Z^0$ at the CERN $\bar p p$ collider in 1983 by teams led by Carlo Rubbia~\cite{WZ} and Pierre Darriulat.
Burt Richter had been on sabbatical at CERN in 1975-1976. During this visit he developed the principles for the scaling of large circular $e^+ e^-$ colliders~\cite{Richter:1976ug}. His work triggered CERN to organise technical studies and a working group to study the physics of such machines, which included three Nobel Prizewinners (Richter himself, Rubbia and Jack Steinberger)~\cite{Camilleri}. Mary Gaillard and I wrote a summary of the theoretical interest in such a collider, emphasizing in particular Higgs boson production, both in association with the $Z$ and also in the decay of the $Z$~\cite{EG76}.

In addition to~\cite{EGN}, there were other early theoretical discussions of searches
for the Higgs boson in $e^+ e^-$ collisions at LEP in~\cite{ECFA,LesHouches}.  There are
three important processes for producing the Higgs boson at an $e^+
e^-$ collider: in addition to $Z^0$ decay - $Z^0 \to H + {\bar
  f}f$~\cite{EG76,Bj}, and production in association with the $Z^0$ - $e^+ e^- \to
Z^0 + H$~\cite{EGN} with the cross-section~\cite{IK,LQT}
\beq
\sigma(e^+ e^- \to Z + H) \; = \; \frac{\pi \alpha^2}{24} \left( \frac{2 p}{\sqrt{s}}\right) \left( \frac{p^2 + 3 m_Z^2}{(s - m_Z^2)^2} \right)
\left( \frac{1 - 4 \sin^2\theta_W + 8 \sin^4 \theta_W}{\sin^2 \theta_W (1 - \sin^2 \theta_W)^2} \right) \, ,
\eeq
where $p$ is the momentum of the final-state particles, there is also production via $W^+ W^-$ or $Z^0 Z^0$ fusion,
$e^+ e^- \to {\bar \nu} H \nu, e^+ H e^-$~\cite{JP}, which becomes important at high energies.  The direct
process $e^+ e^- \to H$ is {\it very} small because - see (\ref{fmass}) - of the
small $H$ coupling to $e^+ e^-$~\cite{EGN}, though detection at FCC-ee cannot be excluded, and the corresponding
reaction at a muon collider, $\mu^+ \mu^- \to H$, may be
interesting~\cite{mumu}.

The LEP project was formally adopted by the CERN Council in 1981. In the early CERN reports on LEP physics the only
discussions of Higgs production were theoretical~\cite{EG76,LesHouches}. The concerns of our
experimental colleagues lay elsewhere, and these early CERN reports contain no experimental
discussions of possible searches for the Higgs boson. The first written experimental discussion of
which I am aware was in an unpublished 1979 report for ECFA~\cite{ECFA}, compiled by a joint working group
of theorists and experimentalists led by Bjorn Wiik. This was followed in 1985 by a more detailed study by a
joint theoretical and experimental working group in a CERN report that Roberto Peccei and I edited, which was published in 1986~\cite{LEP85}.
Thereafter, Higgs searches were firmly in the experimental sights of the LEP collaborations, as seen
in later CERN reports on LEP physics~\cite{sights}.

\begin{figure}[htb]
\begin{center}
\resizebox{6.7cm}{!}{\includegraphics{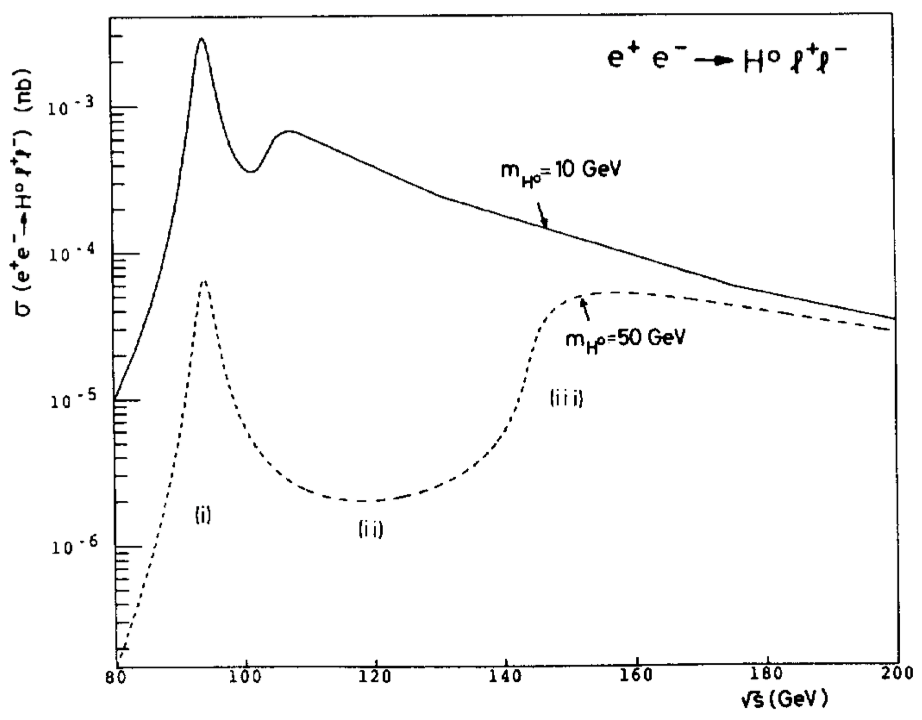}}
\hspace{1cm}
\resizebox{7.3cm}{!}{\includegraphics{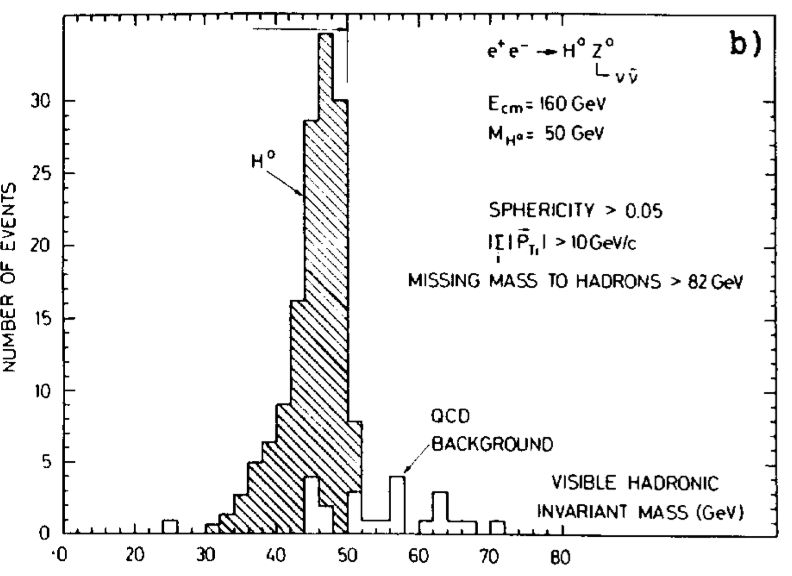}}
\end{center}
\caption{\it Left panel: Cross-sections for the production at LEP of a Higgs boson weighing $10$ or $50$~GeV. Right panel: Simulation of the signal for a $50$-GeV Higgs at LEP produced in a $Z$ boson decaying into $\bar \nu \nu$, compared withe QCD background. Taken from~\cite{LEP85}.}
\label{fig:LEPHiggssearches}
\end{figure}

\begin{figure}[htb]
\begin{center}
\resizebox{7cm}{!}{\includegraphics{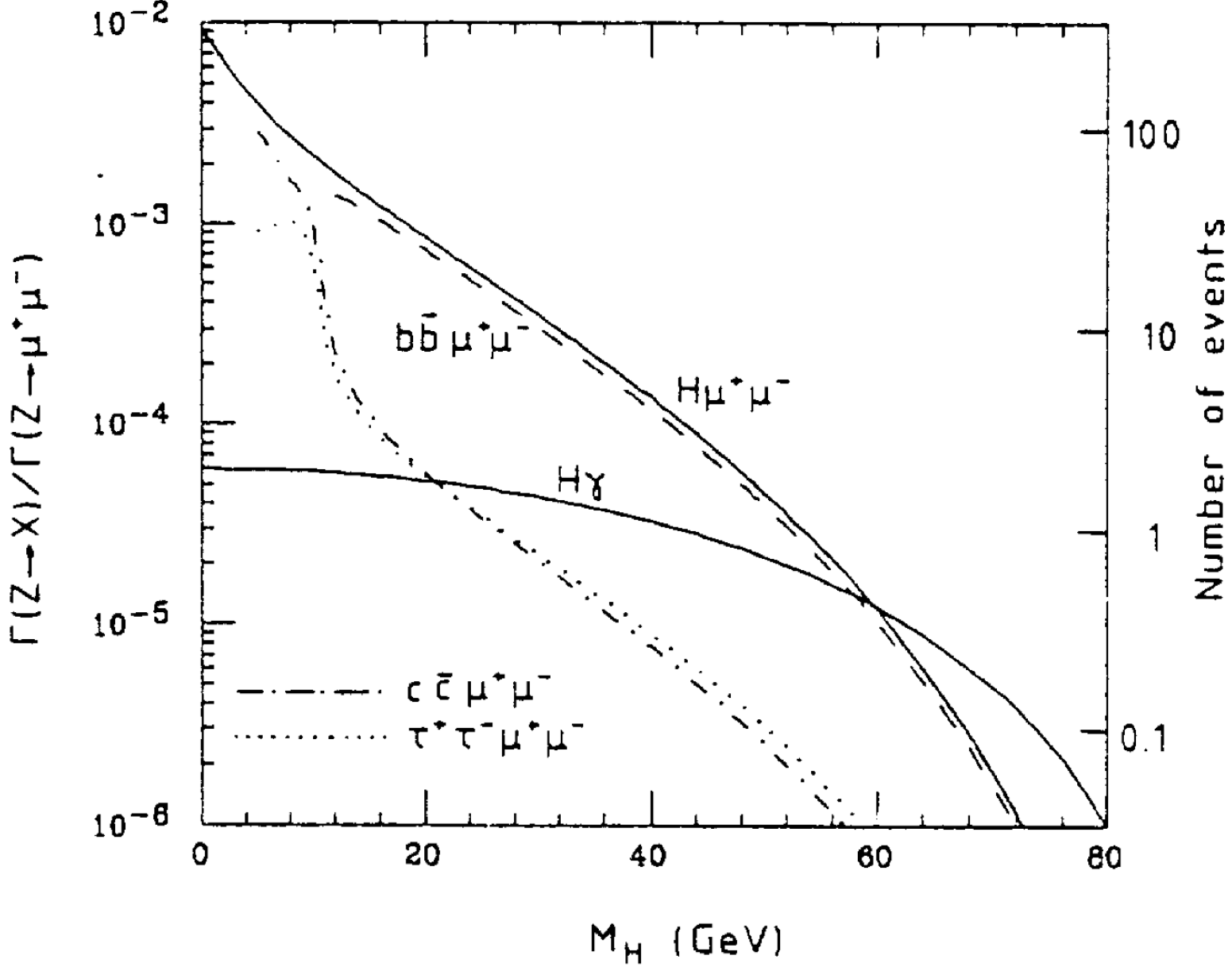}}
\end{center}
\caption{\it Branching ratios for $Z$ decays into a Higgs boson and $\mu^+ \mu^-$ or $\gamma$ compared with $Z \to b \bar b \mu^+ \mu^-, \ c \bar c \mu^+ \mu^-$ and $\tau^+ \tau^- \mu^+ \mu-$, normalised to $Z \to \mu^+ \mu-$ decay. Taken from the second article in~\cite{sights}.}
\label{fig:Franzini+Taxil}
\end{figure}

Figure~\ref{fig:LEPHiggssearches} shows two plots from the 1986 LEP physics study~\cite{LEP85}.
The left panel shows a compilation of the total Higgs production at LEP as a function of the centre-of-mass energy up to 200~GeV for two representative Higgs masses, 10~GeV and 50~GeV, and the right panel shows the result of a simulation of the production of a Higgs boson weighing 50~GeV at a centre-of-mass energy of 160~GeV, showing that it would stand out above the QCD background.
Fig.~\ref{fig:Franzini+Taxil} shows the results of a study of the decays $Z \to H \mu^+ \mu^-$ and $Z \to H \gamma$ compared with background processes $Z \to b \bar b \mu^+ \mu^-, c \bar c \mu^+ \mu^-$ and $\tau^+ \tau^- \mu^+ \mu^-$, taken from the second article in~\cite{sights}.

\begin{figure}[htb]
\begin{center}
\resizebox{7.7cm}{!}{\includegraphics{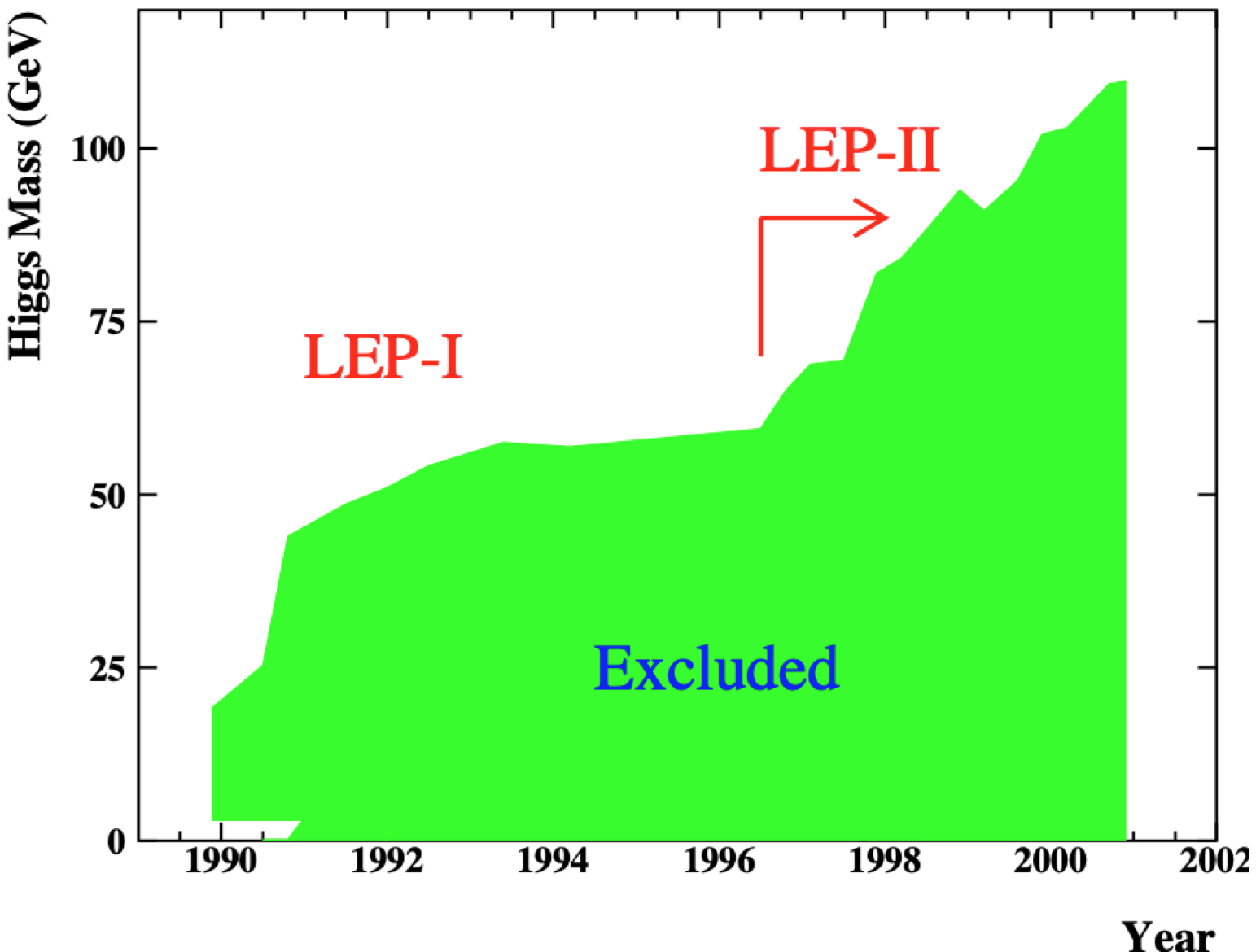}}
\resizebox{7cm}{!}{\includegraphics{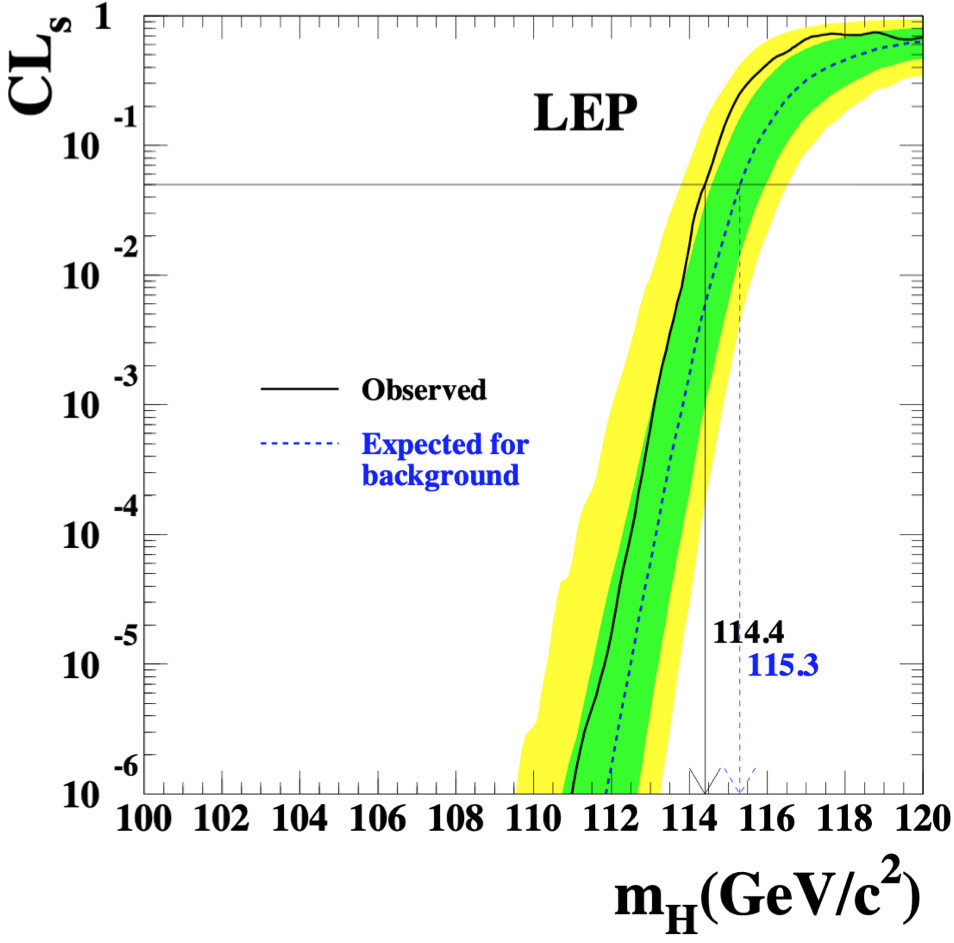}}
\end{center}
\caption{\it Left panel: Evolution of the LEP lower limit on the mass of the Higgs boson~\cite{Mori:2001qow}. Right panel: Final LEP lower limit on the Higgs mass established by the four experimental collaborations~\cite{LEPH}.}
\label{fig:LEP_Higgs_limits}
\end{figure}

Direct searches for the Higgs boson got underway as soon as the LEP experiments started taking data in 1989. 
The non-appearance of the Higgs boson in searches in $Z^0$ decays
required $m_H > 58$~GeV~\cite{LEP1}. As with so many other LEP results, the experimental sensitivity exceeded the previous estimates.
Thereafter, successive increases
in the LEP energy during the LEP~2 era prompted recurrent hopes that
the Higgs discovery might lie just around the corner, but instead the
lower limit on $m_H$ kept rising inexorably~ as shown in the left panel of Fig.~\ref{fig:LEP_Higgs_limits}. Finally, in 2000 the LEP
centre-of-mass energy was pushed to 206~GeV, and a few Higgs-like
events were observed by the ALEPH Collaboration, corresponding to a mass $\sim
115$~GeV~\cite{115}. However, to the disappointment of many physicists, the other LEP experiments did not confirm the ALEPH results, and it was
not possible to push the LEP energy higher. So the difficult decision
was taken to shut LEP down at the end of the year 2000, leaving the
lower limit $m_H > 114.4$~GeV at the 95\%~CL~\cite{LEPH} shown in the right panel of Fig.~\ref{fig:LEP_Higgs_limits}. There was speculation at the time that this decision caused LEP to miss out
on the Higgs discovery. However, with $m_H \approx 125$~GeV as has now
been established~\cite{July4} at the LHC, we now know this was not the case. Substantial extra
investment in accelerating cavities would have been necessary back in
the 1990s in order to be able to push LEP to sufficiently high
energies to produce it, and there had been concerns that the necessary investment would jeopardise the LHC programme.

\section{Constraints on the Higgs Mass from Radiative Corrections}

In parallel with the direct searches for the Higgs boson outlined above, the high-precision
electroweak data obtained at LEP, the SLC and elsewhere made it
possible for the first time to estimate the possible mass of the Higgs boson
via the effects of radiative corrections within the framework of the Standard Model~\cite{EF,EFL}. 
Quantum corrections to electroweak observables had been studied before LEP and the SLC started taking data, and discussed extensively in theoretical reports on LEP physics~\cite{LEP85,sights}. As Tini Veltman~\cite{Veltman}, in particular, pointed out, measurements have a dependence to the top quark mass that is quadratic, whereas the sensitivity to the Higgs mass is logarithmic, e.g., 
\begin{equation}
    m_W^2 \sin^2\theta_W \; = \; m_Z^2 \sin^2\theta_W \cos^2\theta_W \; = \; \frac{\pi \alpha}{\sqrt{2} G_F} (1 + \Delta r)
\end{equation}
where the one-loop radiative correction $\Delta r$ contains contributions
\begin{equation}
    \sim \frac{3 G_F}{8\pi^2 \sqrt{2}} m_t^2, \; \frac{\sqrt{2}G_F}{16 \pi^2}m_W^2 \left( \frac{11}{3} \ln \frac{m_H^2}{m_W^2} \right)
\end{equation}
for large $m_t$ and $m_H$. After LEP and the SLC started their programmes of precision electroweak measurements in 1989 and 1990, much effort went into using them to estimate in advance what the top quark mass might be~\cite{LEPEWWG}. Analyses of indirect constraints predicted a mass that was consistent with the measured value~\cite{Tevatront} within the experimental uncertainties, which were ${\cal O}(10)$~\%.

\begin{figure}[htb]
\begin{center}
\resizebox{6cm}{!}{\includegraphics{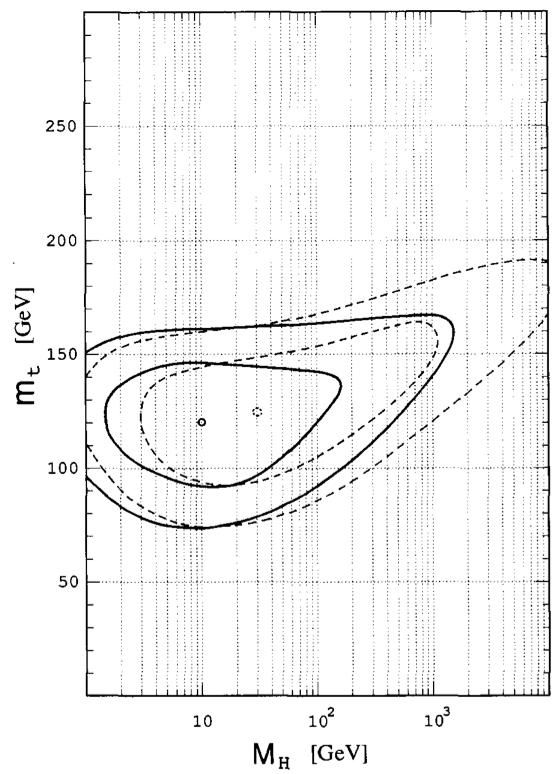}}
\hspace{1cm}
\resizebox{8cm}{!}{\includegraphics{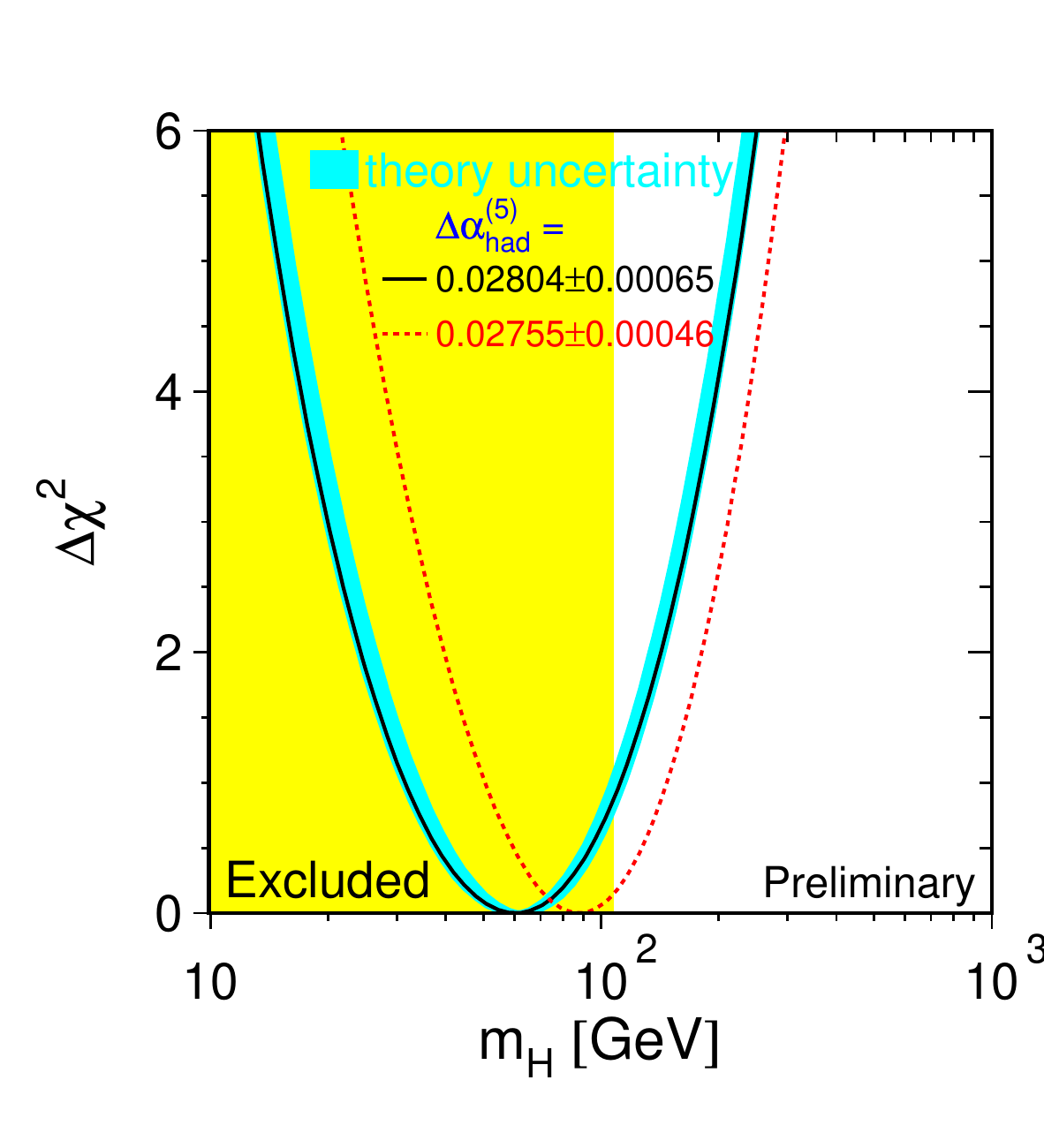}}
\end{center}
\caption{\it Left panel: 68\% and 95\% contours for the Higgs and top quark masses found by analyzing precision electroweak data in~\cite{EFL}. Right panel: Indirect constraint on the mass of the Higgs boson prepared from contributions of the LEP and SLD experiments
to the 2000 Summer conferences~\cite{LEPEWWG2000}.}
\label{fig:LEP_indirect_Higgs_limits}
\end{figure}

The possibility of using the logarithmic $m_H$-dependent
corrections~\cite{Veltman} in conjunction with LEP data to constrain $m_H$ was not discussed
before LEP start-up, perhaps because the corrections were small and the precision of the LEP data had not been anticipated. Indeed, the constraint on $m_H$ was relatively weak before
the top quark was discovered with a mass that agreed with predictions
based on electroweak data, but the measurement of $m_t$ allowed more
accurate estimates of $m_H$ to be made. As seen in the left panel of Fig.~\ref{fig:LEP_indirect_Higgs_limits}, values $< 300$~GeV were favoured
already in the early days of such studies~\cite{EF,EFL},~\footnote{This was an inkling of potential problems with composite Higgs models, whose simplest incarnations~\cite{TC} would have predicted effects similar to those of a rather heavy Higgs.} This indication subsequently matured in 2000 to indicate
that $m_H \sim 100$~GeV~\cite{LEPEWWG2000}, as seen in the right panel of Fig.~\ref{fig:LEP_indirect_Higgs_limits}. 


Something else happened around the same time, namely a calculation of radiative corrections to the mass of the lightest Higgs boson in supersymmetric models, in conjunction with early indications that the top quark might be relatively heavy~\cite{ERZ}. It was found that, whereas the lightest supersymmetric Higgs boson was lighter than the $Z$ boson at the tree level, radiative corrections could push it beyond the $Z$. This was a very expensive calculation, because it suggested that the Higgs could be too heavy to be produced at LEP and you might have to build the LHC. Indeed, calculations indicated that it could weigh 125 GeV, but probably not much heavier than that. Maybe this was just a coincidence, particularly in view of the persistent absence of any direct evidence for supersymmetric particles?

\section{Searches for the Higgs Boson at Hadron Colliders}

The detection of the Higgs boson at hadron colliders is more problematic than
in $e^+ e^-$ collisions. On the one hand, the backgrounds from other physical processes are large,
and on the other hand direct production via the dominant quark constituents in
the proton is small, because they have very small masses~\cite{EGN}. However, loop diagrams (mainly of the top quark) generate a Higgs-gluon-gluon coupling analogous to the $\gamma\gamma$ coupling, as was first pointed out by Frank Wilczek in ~\cite{RU}. As was first discussed in~\cite{GGMN}, this yields a Higgs production cross-section in $pp$ collisions of the form
\beq
\frac{d \sigma}{d y} \; =  \; \left(\frac{\alpha_s}{\pi}\right)^2 \left(\frac{\pi G_\mu}{288 \sqrt{2}}\right) r G(\sqrt{r} e^y) G(\sqrt{r} e^{-y}) \, ,
\label{GGMN}
\eeq
where $y$ is the rapidity, $r \equiv m_H^2/s$ and $G$ is the gluon distribution function within the proton.
This is the dominant Higgs production mechanism at the LHC, and it is ironic that this and
one of the most distinctive Higgs decays, that into $\gamma \gamma$~\cite{EGN}, are both due to
similar quantum effects.
Another important production mechanism is Higgs-strahlung in association with a $W^\pm$ or $Z^0$, 
as was first discussed in~\cite{GNY}, which was the dominant Higgs production mechanism at the Tevatron.
A third important mechanism is $W^+ W^-$ (and $Z^0 Z^0$)
fusion, as first discussed in~\cite{CD}.

A comprehensive theoretical survey of the new physics possibilities at the Superconducting SuperCollider (SSC)
was provided in~\cite{EHLQ}, and the search for the Higgs boson naturally took pride of place. The same
was true in the survey of the new physics possibilities at the LHC provided in~\cite{EGK}, which also
included Higgs production in association with a ${\bar t} t$ pair~\cite{Zoltan}. In anticipation of Higgs 
searches at the SSC, in particular, a comprehensive survey of the theory and phenomenology of the
Higgs boson was published~\cite{HHG}. It served as the Bible for many subsequent Higgs hunters,
also at LEP and the LHC following the much-lamented demise of the SSC.

\begin{figure*}[htb]
\begin{center}
\resizebox{6.9cm}{!}{\includegraphics{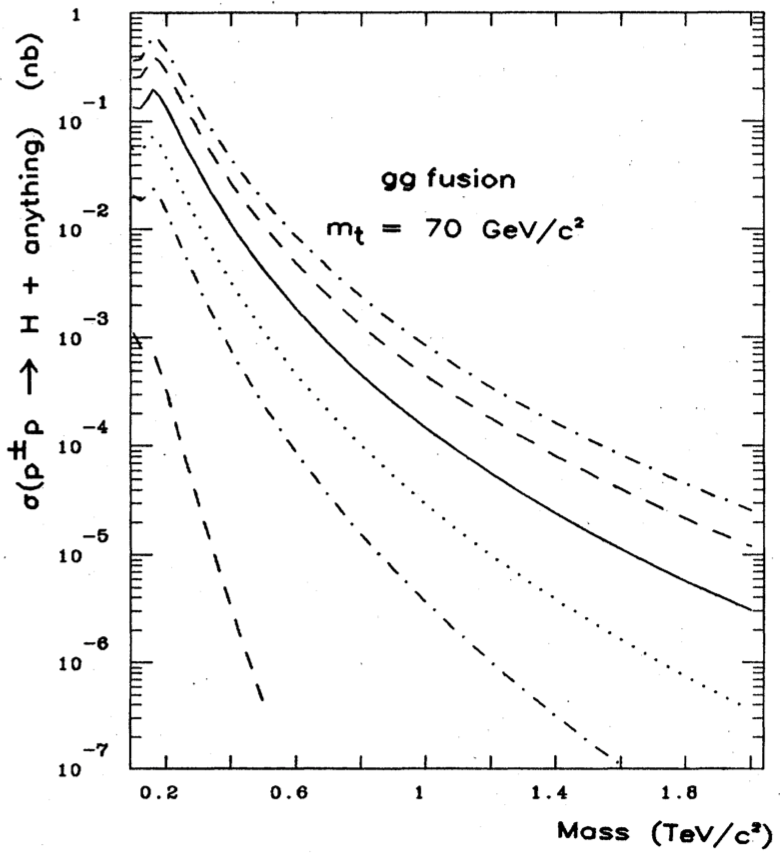}}
\resizebox{7.1cm}{!}{\includegraphics{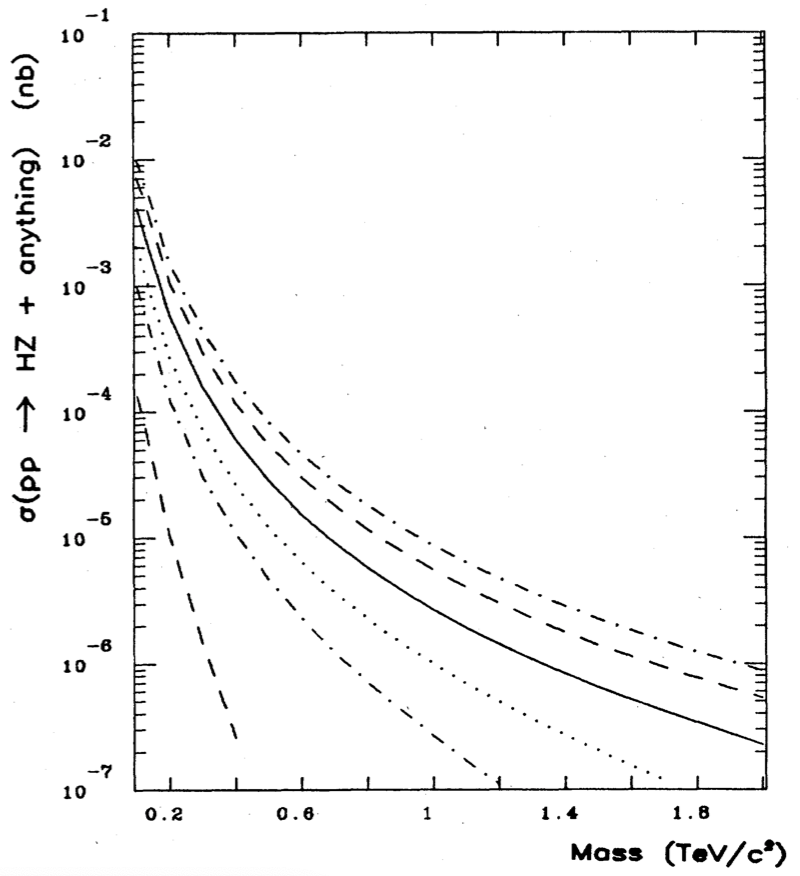}}
\end{center}
\caption{\it Left panel: Cross-sections for Higgs production via gluon-gluon fusion in $p^\pm p$ collisions at $2,10, 20, 40, 70$ and $100$~TeV, assuming $m_t = 70$~GeV. Right panel: Cross-sections for production of the Higgs in association with a $Z$ boson in $pp$ collisions at the same centre-of-mass energies. Taken from~\cite{EHLQ}.}
\label{fig:HiggsXsectionsin pp}
\end{figure*}

\begin{figure*}[htb]
\begin{center}
\resizebox{7cm}{!}{\includegraphics{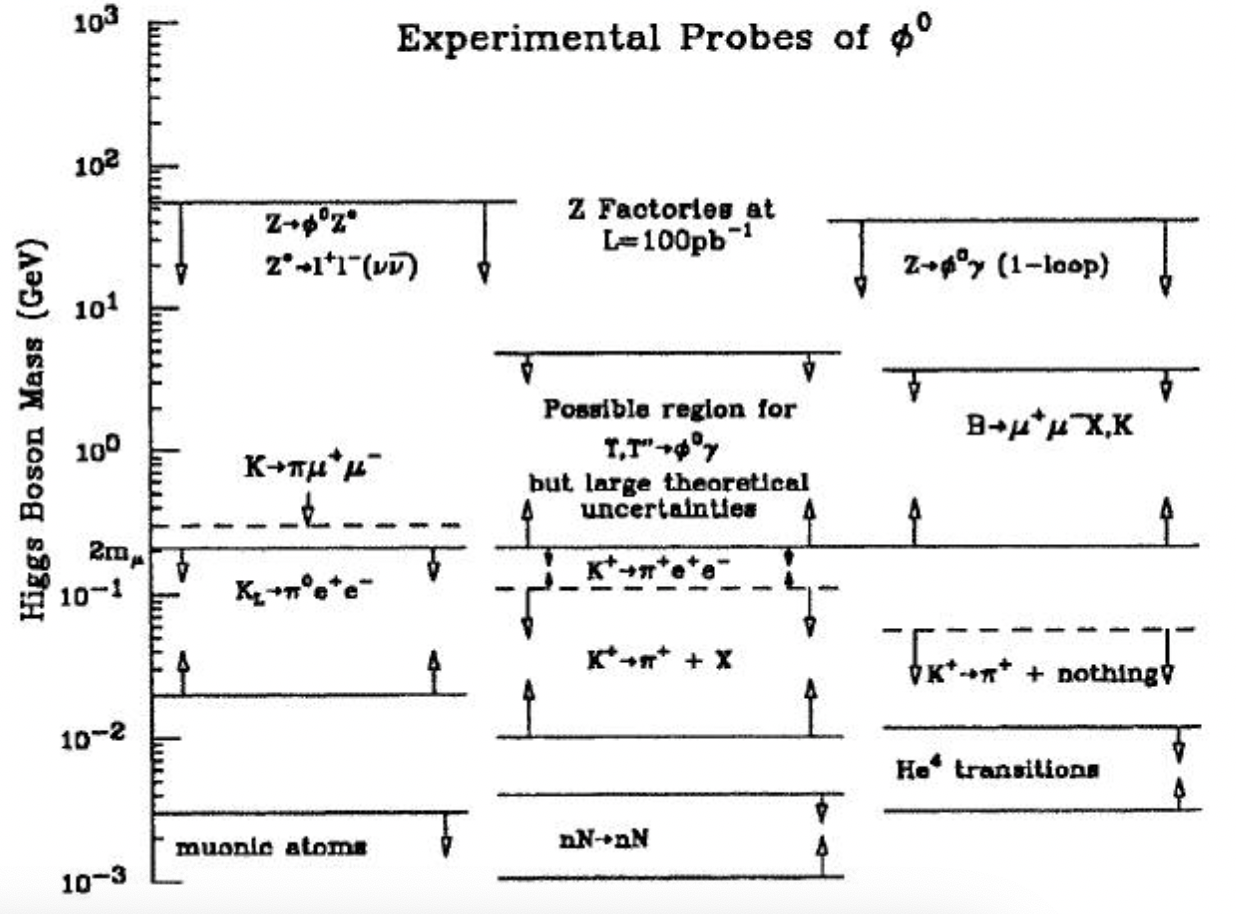}}
\resizebox{7cm}{!}{\includegraphics{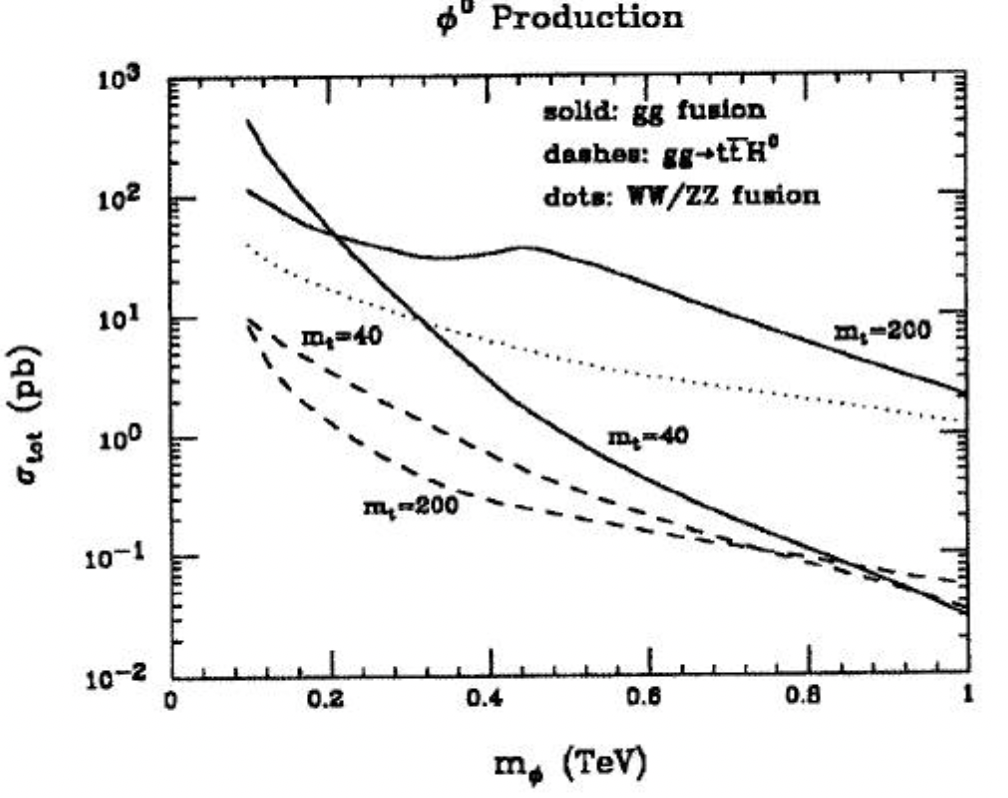}}
\end{center}
\caption{\it Left panel: A compilation of searches for the Higgs boson before LEP. Right panel: Estimates of Higgs production cross-sections in $40$-TeV $pp$ collisions for the indicated values of $m_t$. Taken from~\cite{HHG}.}
\label{fig:HiggsXsectionsatSSC}
\end{figure*}

After the shutdown of LEP, the lead in Higgs searches was taken by the
CDF and D0 experiments at the Tevatron collider, where the dominant
production mechanism was Higgs-strahlung in association with a $W^\pm$
or $Z^0$~\cite{GNY}. As the analyzed Tevatron luminosity accumulated,
CDF and D0 became able to exclude a range of Higgs masses between 156
and 177~GeV~\cite{TeVH}, as well as a range of lower masses in the
range excluded by LEP. There was a small excess of Higgs candidate
events in a range around 130 to 140~GeV, though not enough to
be considered as significant evidence. Since the most
important Higgs decay channels for the Tevatron experiments are $H \to
{\bar b}b$ and $W^+ W^-$, which have relatively poor mass resolution,
the excess did not provide detailed information what value $m_H$ might
have. Unfortunately, the Tevatron was shut down in September 2011,
before it could realise its full potential for Higgs searches.

\begin{figure*}[htb]
\begin{center}
\resizebox{14cm}{!}{\includegraphics{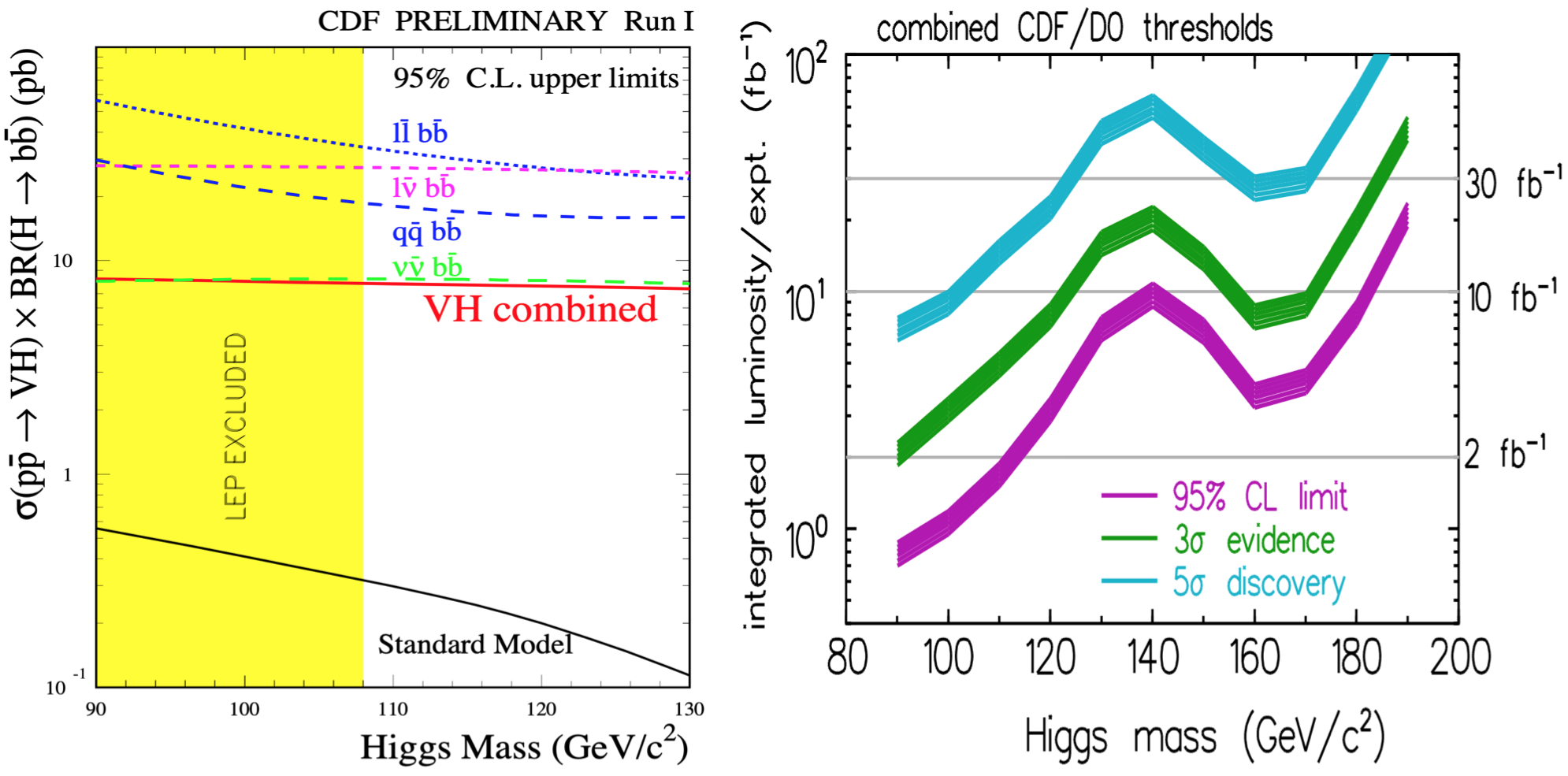}}
\end{center}
\caption{Left panel: Limits from Higgs searches in Run~1 data from the Tevatron~\cite{Valls}, showing also the LEP lower limit on the Higgs mass. Right panel: Estimated Higgs search sensitivity of Tevatron Run~2 data as a function of the assumed integrated luminosity~\cite{HiggsWorkingGroup}.}
\label{fig:TevatronHmasslimits}
\end{figure*}

\section{Afterword}

\begin{figure*}[htb]
\begin{center}
\resizebox{12cm}{!}{\includegraphics{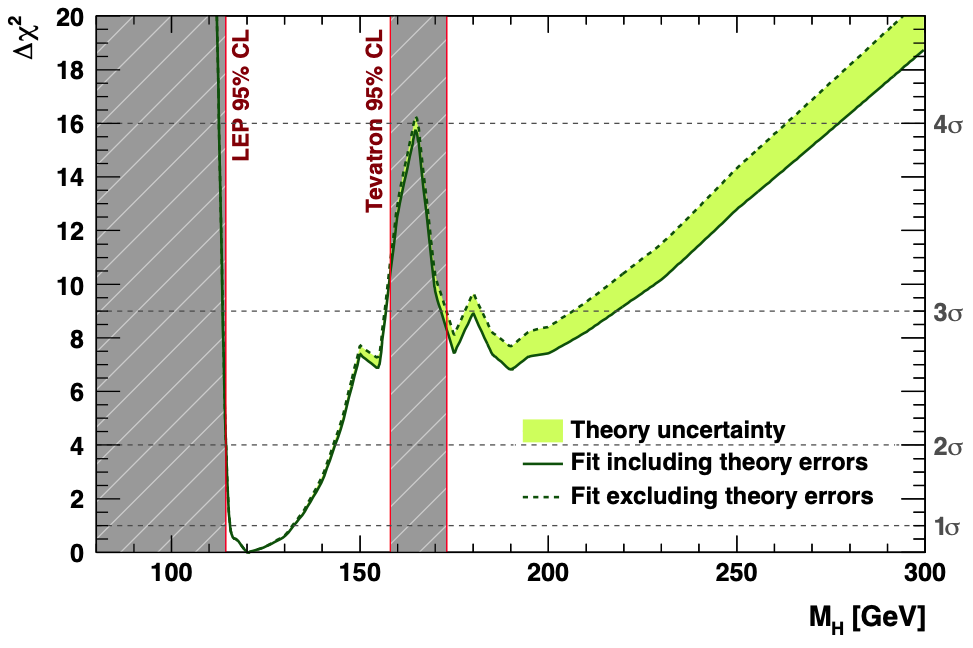}}
\end{center}
\caption{\it A compilation of information about the possible mass of
  the Higgs boson just prior to the LHC discovery~\protect\cite{Gfitter}. The grey shaded regions
  had been excluded by searches at LEP~\protect\cite{LEPH} and the
  Tevatron collider~\protect\cite{TeVH}. The black line
  is the $\chi^2$ function for the combination of these direct searches with precision electroweak data as a
  function of $m_H$, and the green band is its theoretical uncertainty. The dashed
  line is the combination excluding the theoretical uncertainty.  }
\label{fig:Gfitter}
\end{figure*}

I started this talk by discussing events before 1980, and I finish with a few remarks about developments after 2000. Fig.~\ref{fig:Gfitter} is a compilation from the Gfitter group~\cite{Gfitter}, that shows the state of play in 2011, just before the LHC experiments started to close in on the Higgs boson. It shows the exclusion by LEP on the left and the exclusion by the TeVatron experiments in the middle. The curved band is what you get from convoluting those negative searches with the precision and electromagnetic measurements, taking theoretical uncertainties into account. Formally, you find that the best value of the Higgs boson mass is around 125 GeV, but with an uncertainty which is a factor of 100 larger than the current measurement!

\begin{figure*}[htb]
\begin{center}
\resizebox{14cm}{!}{\includegraphics{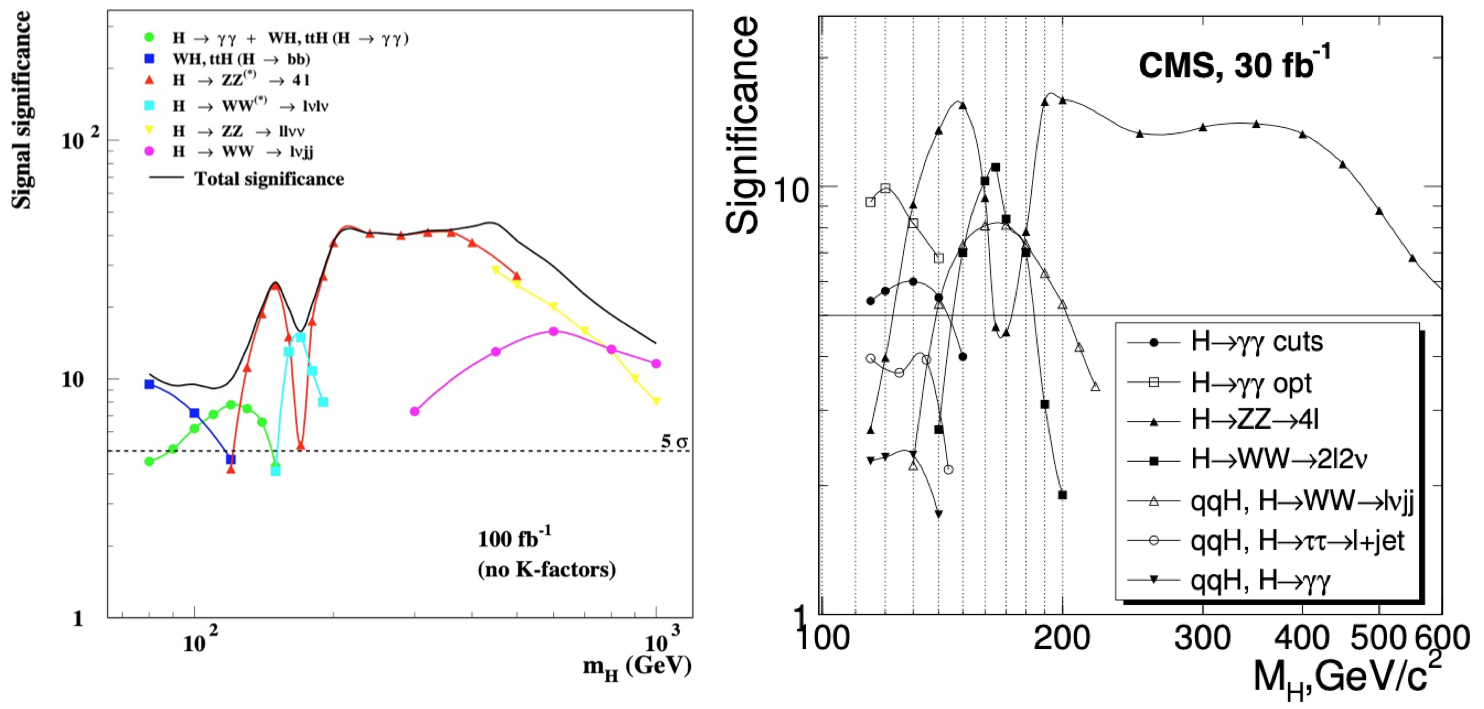}}
\end{center}
\caption{Projected sensitivities of Higgs searches with ATLAS (left panel)~\cite{ATLAS:1999vwa} and CMS (right panel), assuming 100 (30)~fb$^{-1}$ of integrated luminosity~\cite{CMS:2007sch}, respectively.}
\label{fig:LHC_experiments}
\end{figure*}

In parallel to these searches at LEP and the TeVatron, the LHC experiments were busy figuring out how they could discover the Higgs and measure its properties. Fig.~\ref{fig:LHC_experiments} shows a couple of plots. That on the left is taken from the ATLAS Physics TDR published in 1999~\cite{ATLAS:1999vwa} and that on the right is from the CMS Physics TDR~\cite{CMS:2007sch},
which appeared a few years later. ATLAS and CMS guaranteed the discovery of the Higgs boson with 100~fb$^{-1}$, but they discovered it with about 10~fb$^{-1}$~\cite{July4}. Sometimes Nature is kind.

However, between 1980 and 2000, Nature was not so kind. Although she gave us some hints, the Higgs boson did not show up in the period that I was asked to cover.


\section*{Acknowledgements}

This work was supported by the STFC (UK) via the Research Grant ST/T000759/1.


\end{document}